\documentclass[aps,prd,twocolumn,showpacs,floats,floatfix,letterpaper,nofootinbib,superscriptaddress,]{revtex4}

\usepackage{amssymb,amsmath,latexsym,mathrsfs}
\usepackage{graphicx}
\usepackage{epsfig}

\newcommand{\Cc}{{\cal C}}

\newcommand{\CC}{{\bf C}}

\newcommand{\Veff}{V_{\rm eff}}
\newcommand{\Vsur}{V_{\rm survey}}
\newcommand{\kmax}{k_{\rm max}}

\begin{document}


\title{Constraints on massive sterile neutrino species from current and future cosmological data}

\author{Elena Giusarma}
\affiliation{IFIC, Universidad de Valencia-CSIC, 46071, Valencia, Spain}
\author{Martina Corsi}
\affiliation{Physics Department and INFN, Universita' di Roma 
	``La Sapienza'', Ple.\ Aldo Moro 2, 00185, Rome, Italy}
\author{Maria Archidiacono$^2$}
\author{Roland de Putter$^1$}
\author{Alessandro Melchiorri$^2$}
\author{Olga Mena$^1$}
\noaffiliation
\author{Stefania Pandolfi}
\affiliation{Physics Department and ICRA, Universita' di Roma 
	``La Sapienza'', Ple.\ Aldo Moro 2, 00185, Rome, Italy}
\noaffiliation

\begin{abstract}
Sterile massive neutrinos are a natural extension of the Standard Model of elementary particles.
The energy density of the extra sterile massive states affects cosmological measurements in an
analogous way to that of active neutrino species. We perform here an analysis of current cosmological
data and derive bounds on the masses of the active and the sterile neutrino states as well as on the
number of sterile states. The so-called (3+2) models with three sub-eV active massive neutrinos
plus two sub-eV massive sterile species is well within the $95\%$ CL allowed regions when
considering cosmological data only. If the two extra sterile states have thermal abundances at
decoupling, Big Bang Nucleosynthesis bounds compromise the viability of (3+2) models.
Forecasts from future cosmological data on the active and sterile neutrino parameters are
also presented. Independent measurements of the neutrino mass from tritium beta decay
experiments and of the Hubble constant could shed light on sub-eV massive sterile neutrino scenarios.
\end{abstract}


\maketitle

\section{Introduction}

Solar, atmospheric, reactor, and accelerator neutrinos have provided compelling evidence for the existence of neutrino oscillations, implying non-zero neutrino masses (see Ref.~\cite{GonzalezGarcia:2007ib} and references therein). The present data require the number of massive neutrinos to be equal or larger than two, since there are at least two mass squared differences ($\Delta m^2_{\rm atmos}$ and $\Delta m^2_{\rm solar}$) driving the atmospheric and solar neutrino oscillations respectively. Unfortunately, oscillation experiments only provide bounds on the neutrino mass squared differences, i.e. they are not sensitive to the overall neutrino mass scale.

Cosmology provides one of the means to tackle the absolute scale of neutrino
masses. Neutrinos can leave key signatures in several cosmological data sets. 
The amount of primordial relativistic neutrinos changes the epoch of the matter-
radiation equality, leaving an imprint on both Cosmic Microwave Background (CMB) anisotropies (through the so-called Integrated Sachs-Wolfe effect) and on 
structure formation, while non relativistic neutrinos in the recent Universe 
suppress the growth of matter density fluctuations and galaxy clustering, see
Ref.~\cite{Lesgourgues:2006nd}. Cosmology can therefore weigh neutrinos, providing an upper bound on the sum of the three active neutrino masses, $\sum m_\nu \sim 0.58$~eV at the $95\%$ CL~\cite{Komatsu:2010fb}. The former bound is found when CMB measurements from the Wilkinson Microwave Anisotropy Probe (WMAP) are combined with measurements of the distribution of galaxies (SDSSII-BAO) and of the Hubble constant $H_0$ (HST)~\footnote{For other recent analyses, see also Refs.~\cite{Reid:2009nq,Hamann:2010pw}.} in the assumption of a flat universe with a cosmological constant, i.e. a $\Lambda$CDM cosmology.

However, the three neutrino scenario is a minimal scheme, and there is no fundamental symmetry in nature forcing  a definite number of right-handed (sterile) neutrino species, as those are allowed in the Standard Model fermion content. Indeed, cosmological probes have been extensively used to set bounds on the relativistic energy density of the universe in terms of the effective number of neutrinos $N_\nu^{\rm eff}$ (see, for instance, Refs.~\cite{Mangano:2006ur,Hamann:2007pi,Reid:2009nq,Hamann:2010pw,Mangano:2010ei}). Currently,  WMAP, SDSSII-BAO and HST data provide a $68\%$ CL range on $N_\nu^{\rm eff}=4.34^{+0.86}_{-0.88}$~\cite{Komatsu:2010fb} in the assumption of a $\Lambda$CDM universe. If the effective number of neutrinos $N_\nu^{\rm eff}$ is larger than the Standard Model prediction of $N_\nu^{\rm eff}=3.046$ at the Big Bang Nucleosynthesis (BBN) era, the relativistic degrees of freedom, and, consequently, the Hubble expansion rate will also be larger causing weak interactions to become uneffective earlier. This will lead to a larger neutron-to-proton ratio and will change the standard BBN predictions for light element abundances. Combining Deuterium and $^4$He data, the authors of Ref.~\cite{Mangano:2006ur} found $N_\nu^{\rm eff}=3.1^{+1.4}_{-1.2}$ at the $95\%$ CL. 

Models with one additional $\sim 1$~eV massive sterile neutrino, i.e. the so called (3+1) models, were introduced to explain LSND short baseline (SBL) antineutrino data~\cite{Aguilar:2001ty} by means of neutrino oscillations. A much better fit to SBL appearance data and, to a lesser extent, to disappearance data, is provided by models with two sterile neutrinos (3+2)~\cite{Sorel:2003hf,Karagiorgi:2006jf} which can also explain both the MiniBooNE neutrino~\cite{AguilarArevalo:2007it} and antineutrino data~\cite{AguilarArevalo:2009xn} if CP violation is allowed~\cite{Karagiorgi:2009nb}. CP violation can even occur in (3+1) scenarios with only one relevant mass squared difference in the presence of non standard neutrino interactions (NSI). Therefore, the (3+1) NSI model can also nicely explain current data~\cite{Akhmedov:2010vy}. While (3+1) and (3+2) models show some tension with BBN bounds on $N_\nu^{\rm eff}$, the extra sterile neutrinos do not necessarily have to feature thermal abundances at decoupling. The first analysis of both SBL oscillation data and cosmological data was performed by the authors of Ref.~\cite{Melchiorri:2008gq}, where the usual full thermalization scenario for the sterile neutrino species was not assumed. Instead, the sterile abundances were computed taking into account the multi flavour mixing processes operating at the neutrino decoupling period. Robust bounds on sterile neutrino masses, mixings and abundances were derived. However, the masses of the three active neutrinos were fixed to $m_1\sim 0$, $m_2\sim \sqrt{\Delta m^2_{\rm solar}}$ and $m_3 \sim \sqrt{\Delta m^2_{\rm atmos}}$. In Ref.~\cite{Acero:2008rh} the authors derived the bounds on a light sterile neutrino scenario enlarging the usual thermal scenario. More recently, the authors of Ref.~\cite{rt} have used current cosmological data to analyze two possible active plus sterile neutrino scenarios, one with massless active neutrinos (and massive steriles) and the other one with massless steriles states of unknown number (and massive active species). 
However, there are no cosmological bounds on the more natural and oscillation-data motivated scenario in which both the sterile and the active neutrinos have masses. Active neutrinos are \emph{massive}; this is what oscillation data are telling us. In the same way, the LSND and MiniBooNE antineutrino data, if explained in terms of neutrino oscillations, point to the existence of \emph{massive} sterile neutrino species. What oscillation data can not tell us is the absolute scale of neutrino masses and this is precisely what we address in this study, in the spirit of Ref.~\cite{Dodelson:2005tp}, via present and future cosmological measurements.

The paper is organized as follows. In Sec. \ref{sec:i} we present the constraints on the active and sterile neutrino masses and on the number of sterile species from current cosmological data as well as from BBN measurements of light element abundances. Section \ref{sec:ii} is devoted to future errors on these parameters. We describe the Fisher matrix method used here for forecasting errors and discuss the potential results from the ongoing Planck CMB mission combined with future BOSS and Euclid galaxy survey data. We also describe the induced biases on some parameters (such as $H_0$ and $m_\nu$) when the cosmological model does not account for the presence of sterile states to describe the data. We conclude in Sec. \ref{sec:seciii}.

\section{Current constraints}
\label{sec:i}
Here we summarize the constraints from current data on the active neutrino masses and on the sterile neutrino thermal abundance and masses. We have modified the Boltzmann CAMB code~\cite{camb} incorporating the extra massive sterile neutrino parameters and extracted cosmological parameters from current data using a Monte Carlo Markov Chain (MCMC) analysis based on the publicly available MCMC package \texttt{cosmomc}\cite{Lewis:2002ah}. We consider here a flat $\Lambda$CDM scenario plus three ($N_{\nu_s}$) active (sterile) massive neutrino species, described by a set of cosmological parameters
\begin{equation}
 \label{parameter}
  \{\omega_b,\omega_c, \Theta_s, \tau, n_s, \log[10^{10}A_{s}], m_\nu , m_{\nu_s}, N_{\nu_s}\}~,
\end{equation}
where $\omega_b\equiv\Omega_bh^{2}$ and $\omega_c\equiv\Omega_ch^{2}$ are the physical baryon and
cold dark matter densities, $\Theta_{s}$ is the ratio between the sound horizon and the angular diameter
distance at decoupling, $\tau$ is the optical depth, $n_s$ is the scalar spectral index, $A_{s}$ is the
amplitude of the primordial spectrum~\footnote{The pivot scale assumed in this study corresponds to $k_0=0.05$~Mpc$^{-1}$.}, $m_\nu$ is the active neutrino mass, $m_{\nu_s}$ is the sterile neutrino
mass and $N_{\nu_s}$ is the number of thermalized sterile neutrino species. We assume that both active and
sterile neutrinos have degenerate mass spectra ($m_{\nu}$ and $m_{\nu_s}$ are the individual masses, not
the sum of the masses). The flat priors assumed on these cosmological parameters
are shown in Tab.~\ref{tab:priors}.
\begin{table}[htbp]
\begin{center}
\begin{tabular}{c|c}
\hline\hline
 Parameter & Prior\\
\hline
$\Omega_{b}h^2$ & 0.005-0.1\\
$\Omega_{c}h^2$ & 0.01-0.99\\
$\Theta_s$&0.5-10\\
$\tau$ & 0.01-0.8\\
$n_{s}$ & 0.5-1.5\\
$\ln{(10^{10} A_{s})}$ & 2.7-4\\
$m_{\nu_s}$\ [eV] &  0-3\\
$m_{\nu}$\ [eV] &  0-3\\
 $N_{\nu_s}$ &  0-6\\
\hline\hline
\end{tabular}
\caption{Flat priors for the cosmological parameters considered here.}
\label{tab:priors}
\end{center}
\end{table}

Our basic data set is the seven--year WMAP data \cite{Komatsu:2010fb,wmap7}  (temperature and polarization) with the routine for computing the likelihood supplied by the WMAP team. We consider two cases: we first analyze the WMAP data together with the luminous red galaxy clustering results from SDSSII (Sloan Digital Sky Survey)~\cite{beth} and with a prior on the Hubble constant from HST (Hubble Space Telescope)~\cite{Riess:2009pu}, referring to it as the  ``run1'' case. We then include to these data sets Supernova Ia Union Compilation 2 data~\cite{sn}, and we will refer to this case as ``run2''. In addition, we also add to the previous data sets the BBN measurements of the $^4$He abundance, considering separately helium fractions of $Y^1_p=0.2561\pm 0.0108$ (see Ref.~\cite{aver}) and of $Y^2_p=0.2565\pm 0.0010$ (stat.) $\pm 0.0050$ (syst.) from Ref.~\cite{it}. Finally, we also consider the Deuterium abundance measurements $\log (D/H) =−4.56 \pm  0.04$ from Ref.~\cite{pettini}.

It is important to clarify that CMB anisotropies also depend on the value of $Y_p$ but since $Y_p$ is constrained
loosely by current CMB/LSS data, it is consistent to fix it to value $Y_p=0.24$ in the CMB runs and to consider it as an independent parameter constrained by BBN observations.

Given a cosmological model, we predict the theoretical primordial abundance of
$Y_p$ and $\log (D/H)$ by making use of the public available PArthENoPE
BBN code (see \cite{iocco}).

Since running cosmomc and getting at the same time the theoretical predictions from Parthenope 
for BBN would be be exceedingly time-consuming we perform importance sampling obtaining the 
predicted values for $Y_p$ and $\log (D/H)$ with an interpolation routine using a grid of 
Parthenope predictions for each ($\omega_b$, $N_{\nu_s}$), as in \cite{hamann2}.

\begin{table}[htbp]
\begin{center}
\begin{tabular}{c c c c c}
\hline\hline
  Parameter & 68\% CL(r1) & 95\% CL(r1) & 68\% CL (r2)& 95\% CL (r2) \\
     \hline
 $N_{\nu_s}$ &$<2.5$ &$<4.1$  &$<2.0$ &$<3.2$ \\
$m_{\nu}$\ [eV] &$<0.13$ &$<0.30$ &$<0.10$ &$<0.20$\\
$m_{\nu_s}$\ [eV]&$<0.22$&$<0.46$ &$<0.20$ &$<0.50$ \\
\hline\hline
\end{tabular}
\caption{1D marginalized bounds on the active and sterile neutrino parameters using the two combinations of data sets described in the text (r1 refers to ``run 1'' and r2 refers to ``run 2'', respectively).}
\label{tab:bfnu}
\end{center}
\end{table}

\begin{table}[htbp]
\begin{center}
\begin{tabular}{ccccc}
\hline\hline
    &$Y^1_p$~\cite{aver}& $Y^2_p$~\cite{it}& $Y^1_p+D$~\cite{pettini} &$Y^2_p+D$~\cite{pettini} \\
     \hline
 $N_{\nu_s}$ & $<2.3$ &$<1.7$  &$<1.7$ &$<1.4$\\
$m_{\nu}$\ [eV] &$<0.17$ &$<0.15$ &$< 0.15$ & $<0.15$ \\
$m_{\nu_s}$\ [eV]&$<0.62$&$<0.67$ &$<0.69$ &$<0.68$ \\
\hline\hline
\end{tabular}
\caption{1D marginalized 95\% CL bounds on $N_{\nu_s}$, $m_{\nu_s}$ and $m_\nu$ after combining the results of ``run 2'' with those coming from different measurements of BBN light element abundances.}
\label{tab:tab_nnus}
\end{center}
\end{table}

Table~\ref{tab:bfnu} shows the 1D marginalized bounds on $N_{\nu_s}$, $m_{\nu_s}$ and $m_\nu$ arising from the two different analyses performed here on cosmological data sets. The marginalized limits have been computed setting a lower limit of 0 in all the three neutrino parameters here explored. The bounds obtained on the parameters associated to the dark matter candidates considered here are consistent
with those obtained in Ref.~\cite{Melchiorri:2007cd} after taking into account the differences in the thermal
abundances of sterile neutrinos and QCD thermal axions. 
When we marginalize over all the cosmological parameters, see Tab.~\ref{tab:bfnu},  the $95\%$ CL upper bound
for $N_{\nu_s}$ is $4.1$ ($3.2$) using ``run1'' (``run2'') data sets. Therefore, current cosmological data
does not exclude at the $95\%$ CL the existence of $\sim 2$ sterile neutrino species with sub-eV masses plus
three sub-eV active massive neutrinos. 
It would be interesting to further explore if a model with sterile neutrinos is preferred over the model with only three active neutrinos.
The results here are also in very good agreement with those of Ref.~\cite{rt} even if in the former analysis
the two species, i.e. the active and sterile neutrino states, were not considered to be massive at the
same time.  

Table~\ref{tab:tab_nnus} shows the $95\%$ 1D marginalized bounds on $N_{\nu_s}$, $m_{\nu_s}$ and $m_\nu$ arising when different combinations of BBN light element abundances measurements are combined with ``run 2'' results. Note that when measurements of the $^4$He abundance are added to CMB, galaxy clustering and SNIa data, the $95\%$ CL upper limit on $N_{\nu_s}$ is $2.3$ ($1.7$) if $Y^1_p=0.2561\pm 0.0108$ ($Y^2_p=0.2565\pm 0.0010 \pm 0.0050$) is assumed. Since the number of sterile species after adding BBN constraints is smaller than before, the sterile  (active) neutrino masses can get slightly larger (smaller) values, since BBN data is insensitive to the dark matter density in the form of massive neutrinos at late times. The combination of Helium and Deuterium abundance measurements compromises the viability of (3+2) models, leading to  $N_{\nu_s}<1.7-1.4$ at the $95\%$ CL. However, the two sterile states might not have thermal properties at decoupling and evade BBN constraints. A complete analysis~\cite{prep} including sterile neutrino mixing parameters and recent reactor neutrino oscillation results~\cite{Mention:2011rk} is mandatory.

\begin{figure}[h]
\begin{tabular}{c} 
\includegraphics[width=6cm]{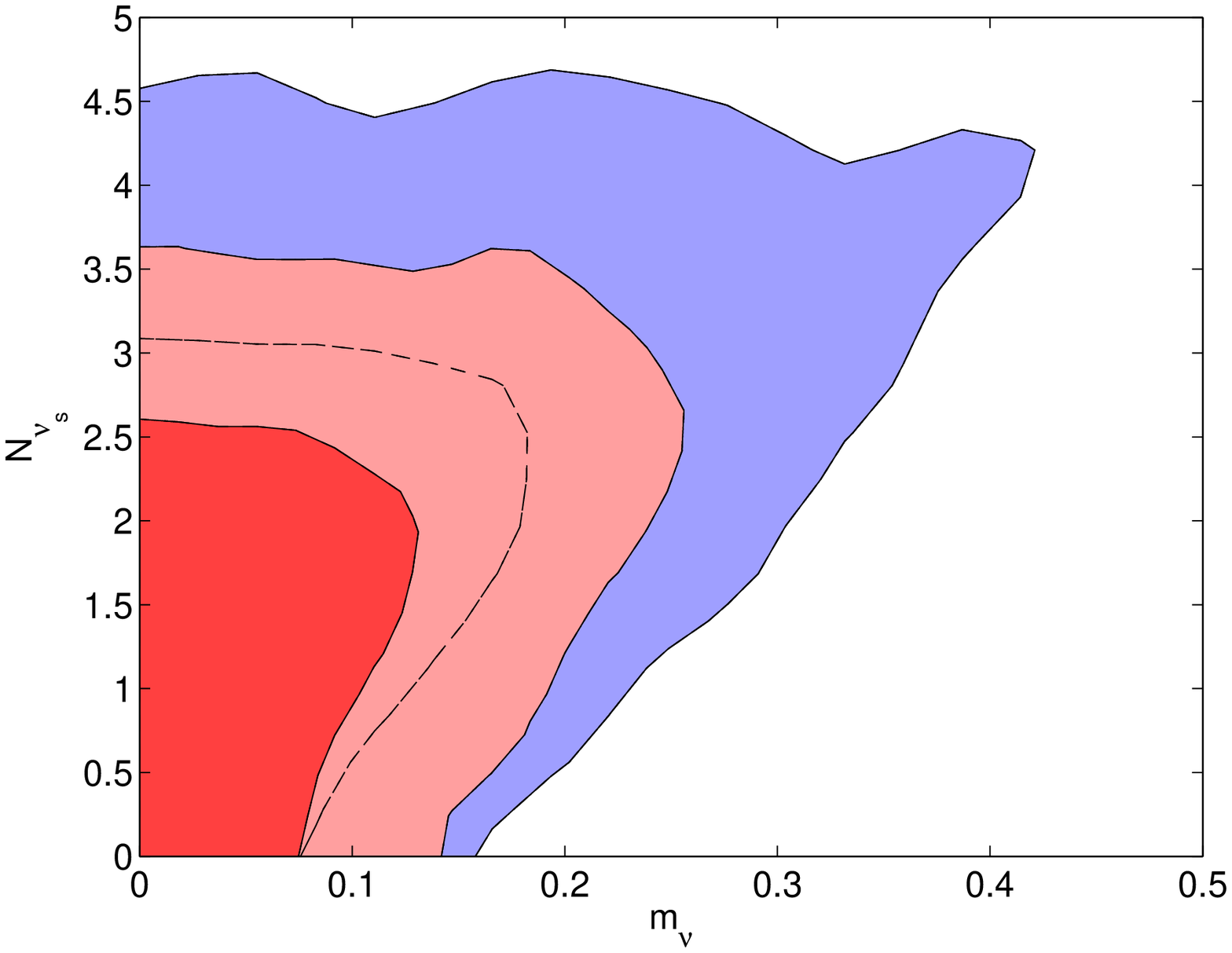}\\
\includegraphics[width=6cm]{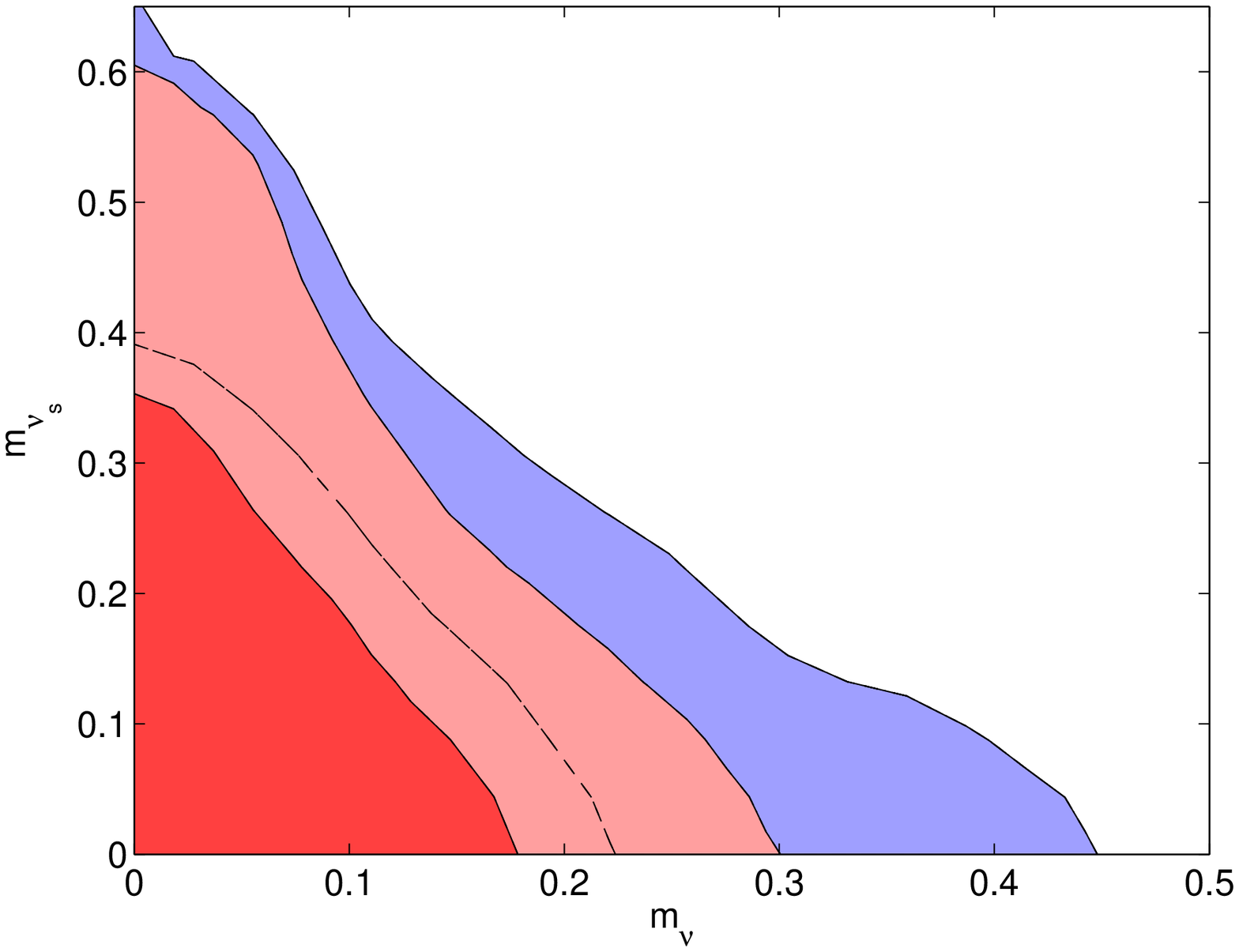}\\
\includegraphics[width=6cm]{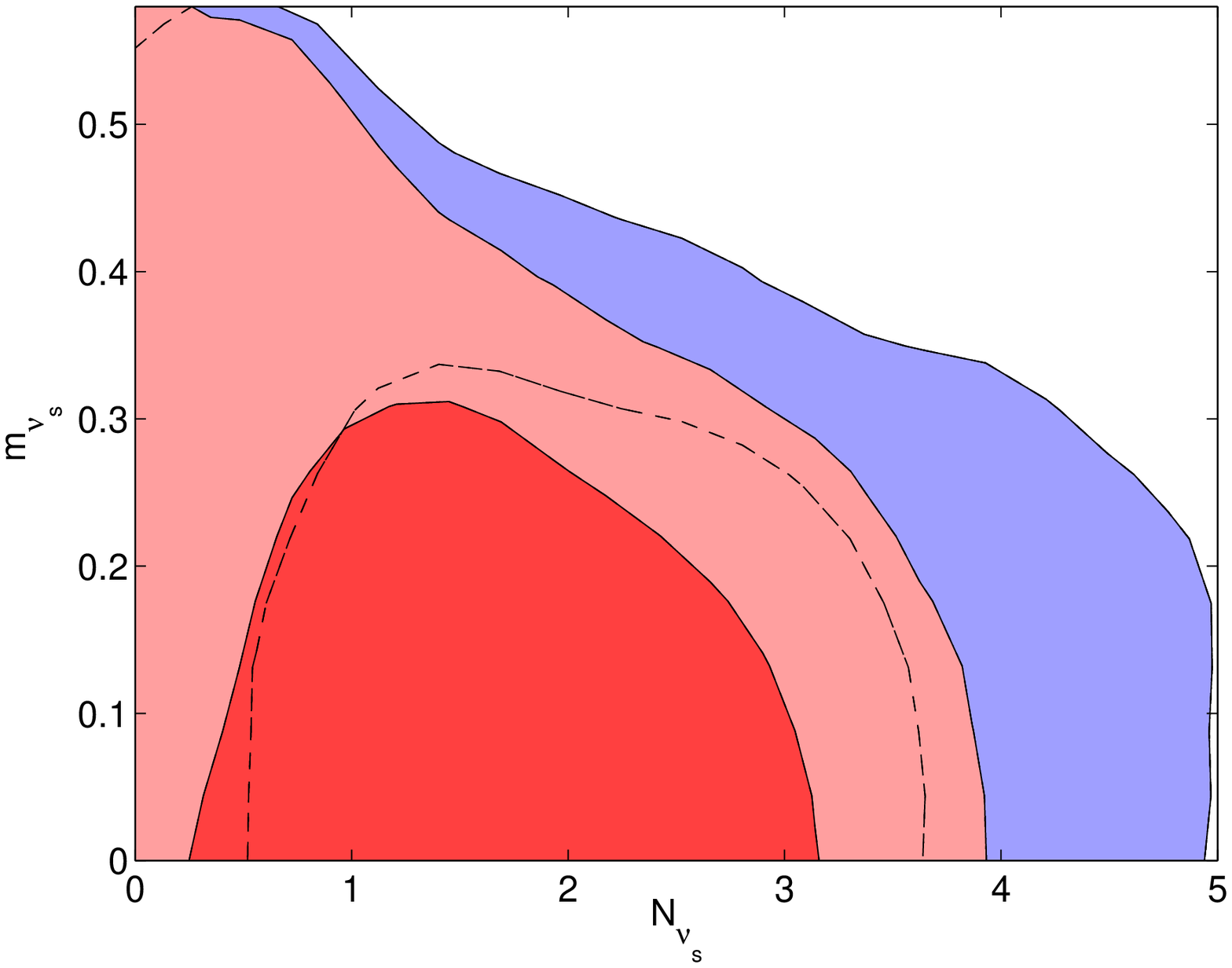}\\
\end{tabular}
\caption{The top, middle and bottom panels show the 68\%  and 95\% CL constraints on the plane $m_{\nu}$-$N_{{\nu}_s}$, $m_{\nu}$-$m_{{\nu}_s}$ and $m_{{\nu}_s}$-$N_{{\nu}_s}$, respectively. The blue (red) contours denote the allowed regions by ``run1'' (``run2'') data sets, see text for details. The masses of the sterile and active neutrinos are both in eV units.}
\label{fig:mnu_mnus}
\end{figure}

Figure~\ref{fig:mnu_mnus}, top panel, depicts the $68\%$ and $95\%$ CL allowed 
contours in the $m_{\nu}$--$N_{{\nu}_s}$ plane. The blue (red) contours denote the allowed regions by ``run1'' (``run2'') data sets. 
Notice that there exists a degeneracy between these two quantities. 
This degeneracy is similar to the one found by the authors of Ref.~\cite{rt}.
When the mass energy density in the form of massive neutrinos is increased, the number of extra relativistic species must also be increased to compensate the effect. This will be the case for massless sterile species. In our analysis, the degeneracy is milder since sterile neutrinos are massive and therefore they behave as an additional dark matter component at late times. The degeneracy will show up when the active neutrinos have relatively large masses, since, in that case, a tiny amount of sterile neutrino masses will be allowed. The sterile states will then behave as relativistic particles at the decoupling era and will compensate the effect of a large active neutrino mass. 

Figure~\ref{fig:mnu_mnus}, middle panel, depicts the $68\%$ and $95\%$ CL 
allowed contours  in the $m_{\nu}$--$m_{{\nu}_s}$ plane. There exists 
a very strong anticorrelation between these two quantities, since both 
contribute to the dark matter energy density at late times
and therefore if the mass of the sterile neutrino states grows, the mass of 
the active ones must decrease. The situation is analogous to that 
of QCD thermal axions and massive (active) neutrinos, see Ref.~\cite{Melchiorri:2007cd}. 

The bottom panel of Fig.~\ref{fig:mnu_mnus} depicts the $68\%$ and $95\%$ CL allowed contours in the $N_{\nu_s}$--$m_{{\nu}_s}$ plane. In this case, the larger the sterile neutrino mass is, the lower its thermal abundance must be, as expected. 

\section{Future constraints}
\label{sec:ii}
We present here the constraints on the neutrino sector parameters explored in this work from future CMB and galaxy survey measurements, making use of the Fisher matrix formalism, see also Ref.~\cite{Carbone:2010ik} for a recent analysis.
We also compute the potential shifts in the different cosmological parameters when the sterile neutrino parameters are neglected in the analysis.
\subsection{Methodology}
The Fisher matrix is defined as the expectation value of the second derivative of the likelihood surface about the maximum. As long as the posterior distribution for the parameters is well approximated by a multivariate Gaussian function, its elements are given by~\cite{Tegmark:1996bz,Jungman:1995bz,Fisher:1935bi}
\begin{equation}
\label{eq:fish}
F_{\alpha\beta}=\frac{1}{2}{\rm
  Tr}\left[C^{-1}C_{,\alpha}C^{-1}C_{,\beta}\right]~,
\end{equation}
where $C=S+N$ is the total covariance which consists of signal
$S$ and noise $N$ terms. The commas in Eq.~(\ref{eq:fish})
denote derivatives with respect to the cosmological parameters
within the assumed fiducial cosmology. Our fiducial model is a $\Lambda$CDM  
cosmology with five parameters: the physical baryon and CDM densities, $\omega_b=\Omega_bh^2$ and $\omega_c=\Omega_ch^2$, the scalar spectral index, $n_{s}$, $h$ (being the Hubble constant $H_0=100\ h$~km Mpc$^{-1}$s$^{-1}$) and the dimensionless amplitude of the primordial curvature perturbations, $A_{s}$ (see Tab.~\ref{tab:fiducial_standard_model} for their values). Furthermore, we add to the $\Lambda$CDM fiducial cosmology three additional parameters for the neutrino sector: the mass of active neutrinos $m_\nu$, the mass of sterile neutrinos $m_{\nu_s}$ and the number of sterile neutrino species $N_{\nu_s}$. Notice that, for simplicity, we have kept fixed the reionization optical depth $\tau$ since it has no impact on large scale structure data and we do not expect a strong degeneracy between $\tau$ and the neutrino parameters, see Ref.~\cite{Perotto:2006rj}. We assume that both active and sterile neutrinos have a degenerate spectrum and that the sterile species are fully thermalized. The fiducial values of the neutrino parameters are listed as well in Tab.~\ref{tab:fiducial_standard_model}, and they are based on the constraints from current data presented in the previous section, from which we conclude that $m_\nu=0.1$~eV, $m_{\nu_s}\le 0.5$ and $N_{\nu_s}=1,2$ are within the allowed regions for these parameters.

\begin{table}[htbp]
	\centering
		\begin{tabular}{c|c|c|c|c|c|c|c}
		\hline\hline
    $\Omega_bh^2$ & $\Omega_ch^2$ & $n_{s}$ & $h$ & $A_{s}$&$m_\nu$\ [eV]  &$m_{\nu_s}$\ [eV] &$N_{\nu_s}$\\
		
		0.02267&0.1131&0.96&0.705&$2.64\cdot 10^{-9}$&0.1&0.1-0.5&1-2\\
		\hline\hline
		\end{tabular}
		\caption{Values of the parameters in the fiducial models explored in this study.}
	\label{tab:fiducial_standard_model}
\end{table}

We compute the CMB Fisher matrix to obtain forecasts for the Planck satellite~\cite{Planck}. We follow here the method of Ref.~\cite{Licia05}, considering the likelihood function for a realistic experiment with partial sky coverage, and noisy data
{\small
\begin{eqnarray}
&&-2\ln{\cal L}=\sum_{\ell} (2\ell+1)\Bigg\{f_{sky}^{BB}\ln\left(\frac{\CC_{\ell}^{BB}}{\hat{\CC}_{\ell}^{BB}}\right)+ \nonumber\\
&&+\sqrt{f_{sky}^{TT}f_{sky}^{EE}}\ln\left(\frac{\CC_{\ell}^{TT}\CC_{\ell}^{EE}-
(\CC_{\ell}^{TE})^2}{\hat{\CC}_{\ell}^{TT}\hat{\CC}_{\ell}^{EE}-
(\hat{\CC}_{\ell}^{TE})^2}\right) \nonumber\\
&&+\sqrt{f_{sky}^{TT}f_{sky}^{EE}}\frac{\hat{\CC}_{\ell}^{TT}\CC_{\ell}^{EE}+
\CC_{\ell}^{TT}\hat{\CC}_{\ell}^{EE}-
2\hat{\CC}_{\ell}^{TE}\CC_{\ell}^{TE}}{\CC_{\ell}^{TT}\CC_{\ell}^{EE}-
(\CC_{\ell}^{TE})^2}+\nonumber\\
&&+f_{sky}^{BB}\frac{\hat{\CC}_{\ell}^{BB}}{\CC_{\ell}^{BB}}-2\sqrt{f_{sky}^{TT}f_{sky}^{EE}}-f_{sky}^{BB}\Bigg\}~,
\label{eq:like_real}
\end{eqnarray}}

\noindent
and compute its second derivatives to obtain the corresponding Fisher matrix
\begin{equation}
F^{\rm CMB}_{\alpha\beta}=\left \langle -\frac{\partial^2 {\cal L}}{\partial p_\alpha \partial p_\beta}
\right \rangle|_{{\bf p}=\bar{{\bf p}}} \, .
\label{standard_fish}
\end{equation}
In Eq.~(\ref{eq:like_real}) $\CC^{XY}_{\ell}=\Cc^{XY}_{\ell}+{\cal
N}^{XY}_{\ell}$ with $\Cc^{XY}_{\ell}$ the temperature and
polarization power spectra ($X,Y \equiv \{T,E,B\}$) and 
${\cal N}_{\ell}$ the noise bias. Finally, $f_{sky}^{XY}$ is the
fraction of observed sky which can be different for the $T$-, $E$-,
and $B$-modes.

For the galaxy redshift survey Fisher matrix, we follow the prescription of Ref.~\cite{eisenstein}. Assuming the likelihood function for the band powers of a galaxy redshift survey to be Gaussian, the Fisher matrix can be approximated as:
\begin{eqnarray}  
F^{\rm LSS}_{\alpha \beta}&=&\int_{\vec{k}_{\rm min}} ^ {\vec{k}_{\rm max}} \frac{\partial \ln P_{\rm gg}(\vec{k})}{\partial p_\alpha} \frac{\partial \ln P_{\rm gg}(\vec{k})}{\partial p_\beta} \Veff(\vec{k}) \frac{d\vec{k}}{2(2 \pi)^3}  
\label{eq:Fij} \\
&=&\int_{-1}^{1} \int_{k_{\rm min}}^{\kmax}\frac{\partial \ln P_{\rm gg}(k,\mu)}{\partial p_\alpha} \frac{\partial \ln P_{\rm gg}(k,\mu)}{\partial p_\beta} \Veff(k,\mu)\nonumber\\
& &\frac{2\pi k^2 dk d\mu}{2(2 \pi)^3}~,  \nonumber 
\end{eqnarray}
where $V_{\rm eff}$ is the effective volume of the survey:
\begin{eqnarray}
\Veff(k,\mu) &=&\left [ \frac{{n}P(k,\mu)}{{n}P(k,\mu)+1} \right ]^2 \Vsur,
\label{eq:Veff} 
\end{eqnarray}
$\mu$ being the cosine of the angle between the vector along the line of sight and $\vec{k}$ and
$n$ being the galaxy number density, which is assumed to be constant throughout the survey.
The linear redshift-space galaxy power spectrum $P_{\rm gg}$ is related to the real-space linear power dark matter spectrum $P_{\rm dm}$ as
\begin{equation}
P_{\rm gg}(k)=P_{\rm dm}(k)(b+f\mu^2)^2
\label{eq:pspectrum}
\end{equation} 
where $b$ is the bias relating galaxy to dark matter overdensities in real space and $f$ is the linear growth factor. Both the bias and the growth factor are assumed to vary in each redshift bin and are considered as additional parameters in the Fisher analysis of galaxy survey data.

We consider here two redshift surveys: the BOSS (Baryon Oscillation Spectroscopic Survey)~\cite{boss} and the Euclid~\cite{Euclid1,Euclid2} experiments. For the BOSS survey we assume a sky area of $10000$~deg$^2$, a redshift range of $0.15 <z <0.65$ and a mean galaxy density of $2.66 \times 10^{-4}$. For Euclid we consider an area of $20000$~deg$^2$, a redshift range of $0.15 <z <1.95$ and a mean galaxy density of $1.56 \times 10^{-3}$. We divide the surveys in redshift bins of width $\Delta{z} = 0.1$ (a value that is much larger than standard redshift spectroscopic errors), set $k_{\rm max}$ to be $0.1h$/Mpc and $k_{\rm min}$ to be greater than $2\pi/\Delta V^{1/3}$, where $\Delta V$ is the volume of the redshift shell. 

Combining the Planck and redshift survey Fisher matrices ($F_{\alpha\beta} =F^{\rm LSS}_{\alpha\beta}+F^{\rm CMB}_{\alpha\beta}$) we get the joint constraints for $\Omega_bh^2$, $\Omega_ch^2$, $n_{s}$, $H_0$, $A_{s}$, $m_\nu$, $m_{\nu_s}$ and $N_{\nu_s}$, after marginalizing over the bias $b$ and the growth factor $f$.  The 1--$\sigma$ error on parameter $p_\alpha$ marginalized over the other parameters is $\sigma(p_\alpha)=\sqrt{({F}^{-1})_{\alpha\alpha}}$,
${F}^{-1}$ being the inverse of the Fisher matrix. 

\subsection{Results}
Tables~\ref{tab:0.31} and \ref{tab:0.32} contain the 1--$\sigma$ marginalized forecasted errors on the cosmological parameters for a fiducial cosmology with $m_\nu=0.1$~eV, $m_{\nu_s}=0.3$~eV and $N_{\nu_s}=1$ and $2$, respectively. We illustrate the results of our Fisher analysis for both BOSS and Euclid galaxy redshift survey data combined with Planck CMB measurements. Note that the errors on the pure $\Lambda$CDM model parameters, i.e $\Omega_bh^2$, $\Omega_ch^2$, $n_{s}$, $h$ and $A_{s}$ are always around or below the $1\%$ level. The error on the active neutrino mass is around $60\%$ for BOSS and half for Euclid. The error on the number of sterile neutrino species is always smaller than $25\%$. Regarding the error on the sterile neutrino mass, it can reach $100\%$ relative errors for BOSS plus Planck data. Naively, one would expect that the BOSS and Euclid errors are related by a factor of $\sqrt{V_{\rm BOSS}/ V_{\rm Euclid}}$ (being $V$ the volume of the survey) when the shot noise is subdominant. However, in practice, the forecasted errors on the pure $\Lambda$CDM parameters are sometimes similar for the BOSS and Euclid cases, which implies that those parameters are mainly determined by CMB measurements. Of course this is not the case for the active and sterile neutrino masses, whose errors are mainly driven by galaxy clustering data and differ by a factor of $\sim 2-3$ for BOSS and Euclid cases, as naively expected. A word of caution is needed here: while computing the errors on the active and sterile neutrino masses and on the sterile neutrino abundances, a $\Lambda$CDM scenario has been chosen as fiducial cosmology. These errors can change if the equation of state of the dark energy component is allowed to vary~\cite{Hannestad:2005gj} and/or interactions between the dark matter and dark energy sectors are stwiched on~\cite{LaVacca:2008mh,Gavela:2009cy}.

We also present here the joint constraints in a two-parameter 
subspace (marginalized over all other cosmological parameters) to study 
the covariance between the sterile neutrino masses and/or abundances and 
the other cosmological parameters considered in this work.  
We have explored several possible scenarios with different sterile neutrino 
masses and thermal abundances (see Tab.~\ref{tab:fiducial_standard_model}). 
However, for the sake of simplicity, we illustrate here only the case $N_{\nu_s}= 1$, $m_{\nu_s}=0.3$~eV and $m_\nu=0.1$~eV. 

\begin{figure}[h]
\begin{tabular}{c c} 
\includegraphics[width=4.3cm]{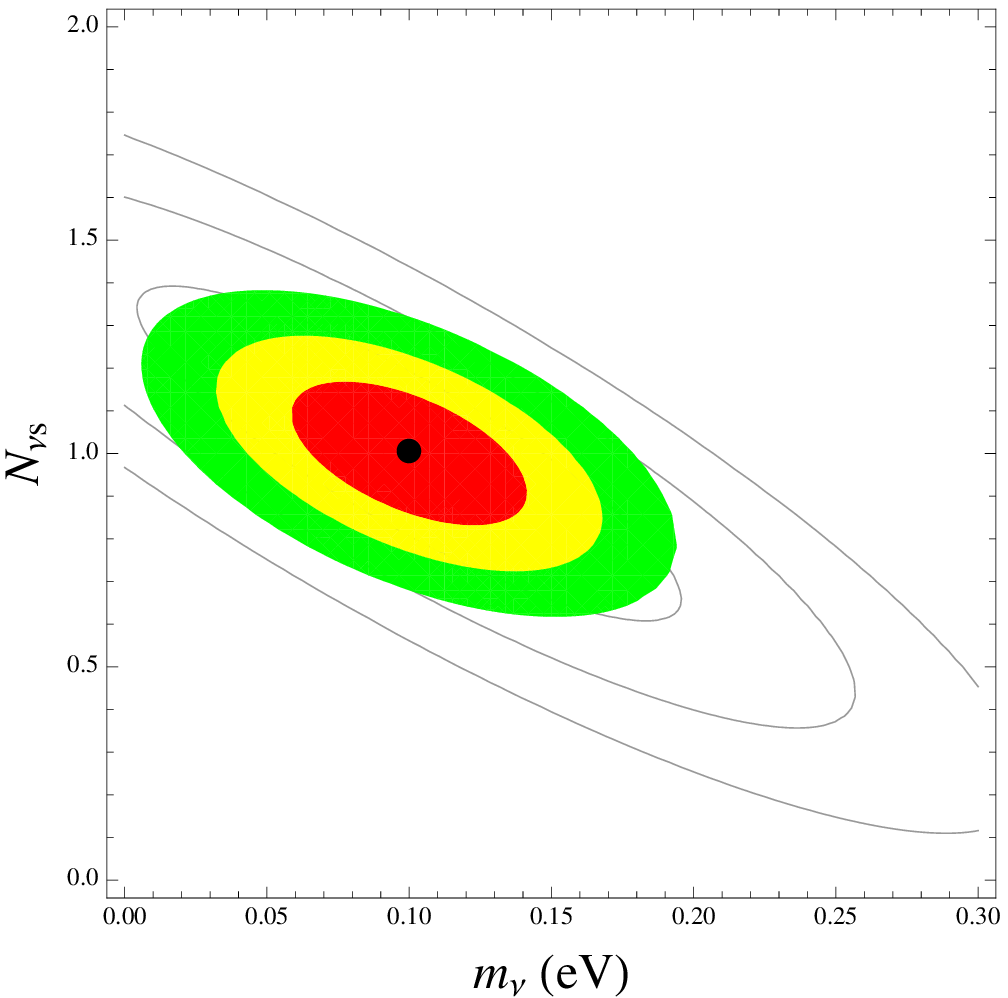}&
\includegraphics[width=4.3cm]{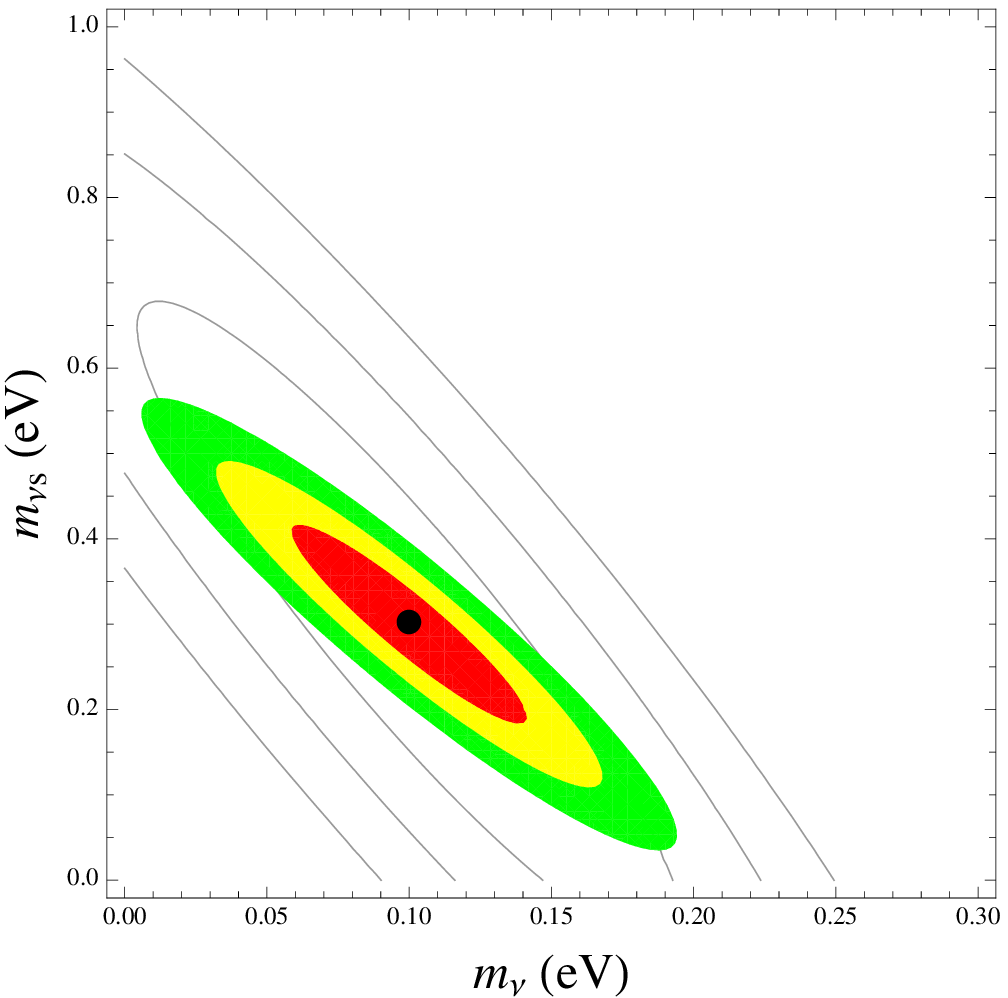}\\
\end{tabular}
\caption{The empty (filled) contours denote the $68\%$, $95\%$ and $99.73\%$ CL regions for Planck plus BOSS (Euclid) data. The neutrino parameters in the fiducial model  are $N_{\nu_s}= 1$, $m_{\nu_s}=0.3$~eV and $m_\nu=0.1$~eV.}
\label{fig_mnu}
\end{figure}

Figure \ref{fig_mnu}, left panel, shows the correlation between the number of sterile species $N_{\nu_s}$ and the active neutrino mass $m_{\nu}$. The expected error on the number of sterile species is very similar for BOSS and Euclid data, which indicates that the constraints on $N_{\nu_s}$ arise mostly from Planck CMB measurements. Since the total energy density in the form of massive neutrinos is the sum of the active plus sterile contributions, a higher neutrino mass is compensated with a lower abundance of massive sterile species. The 1--$\sigma$ marginalized error on $N_{\nu_s}$ from Planck plus BOSS (Euclid) data is 0.26 (0.1), see Tab.~\ref{tab:0.31}. The right panel of Fig.~\ref{fig_mnu} shows the correlation between the masses of sterile and active neutrinos. As expected from the results presented in Fig.~\ref{fig:mnu_mnus} (middle panel) and as previously explained, higher active neutrino masses are allowed for very low values of the sterile neutrino masses. The 1--$\sigma$ marginalized errors on the massive species $m_{\nu_s}$ and $m_{\nu}$ from Planck plus BOSS (Euclid) data are $0.25$ ($0.08$)~eV and $0.06$ ($0.03$)~eV respectively.  If nature has chosen an active neutrino with mass $\sim 0.1$~eV, BOSS (Euclid) data, combined with CMB Planck measurements, could provide a 1.5--$\sigma$ (3--$\sigma$) detection, even in the presence of massive sterile species.   
\begin{table}[htbp]
\begin{center}
\begin{tabular}{cccc}
\hline\hline
 Parameter   &   BOSS+PLANCK   &   EUCLID+PLANCK\\
\hline
$\Omega_{b}h^2$   &   0.7\%   &   0.3\%\\
$\Omega_{c}h^2$   &   2.9\%   &   1.3\%\\
$\ln{(10^{10} A_{s})}$   &   0.7\%   &   0.4\%\\
$h$\ [km/s/Mpc]   &   1.4\%   &   0.7\%\\
$n_{s}$   &   0.6\%   &   0.3\%\\
$m_{\nu}$\ [eV]   &   63.1\%   &   28.0\%\\
$m_{\nu_s}$\ [eV]   &   83.2\%   &   26.2\%\\
$N_{\nu_s}$   &   25.9\%   &   10.6\%\\
\hline\hline
\end{tabular}
\caption{1$\sigma$ marginalized relative errors  for a fiducial cosmology with $N_{\nu_s}$=1, $m_{\nu}$=0.1 eV and $m_{\nu_s}$=0.3 eV.}
\label{tab:0.31}
\end{center}
\end{table}

\begin{table}[htbp]
\begin{center}
\begin{tabular}{ccc}
\hline\hline
Parameter   &   BOSS+PLANCK   &   EUCLID+PLANCK\\
\hline
$\Omega_{b}h^2$   &   0.7\%   &   0.3\%\\
$\Omega_{c}h^2$   &   1.5\%   &   1.7\%\\
$\ln{(10^{10} A_{s})}$   &   0.4\%   &   0.4\%\\
$h$\ [km/s/Mpc]   &   1.4\%   &   0.8\%\\
$n_{s}$   &   0.5\%   &   0.4\%\\
$m_{\nu}$\ [eV]   &   64.9\%   &   35.9\%\\
$m_{\nu_s}$\ [eV]   &   41.9\%   &   16.4\%\\
$N_{\nu_s}$   &   10.2\%   &   7.5\%\\
\hline\hline
\end{tabular}
\caption{1--$\sigma$ marginalized relative errors for all parameters for a $N_{\nu_s}$=2, $m_{\nu}$=0.1 eV and $m_{\nu_s}$=0.3 eV fiducial cosmology.}
\label{tab:0.32}
\end{center}
\end{table}

\begin{figure}[h]
\begin{tabular}{c c} 
\includegraphics[width=4.3cm]{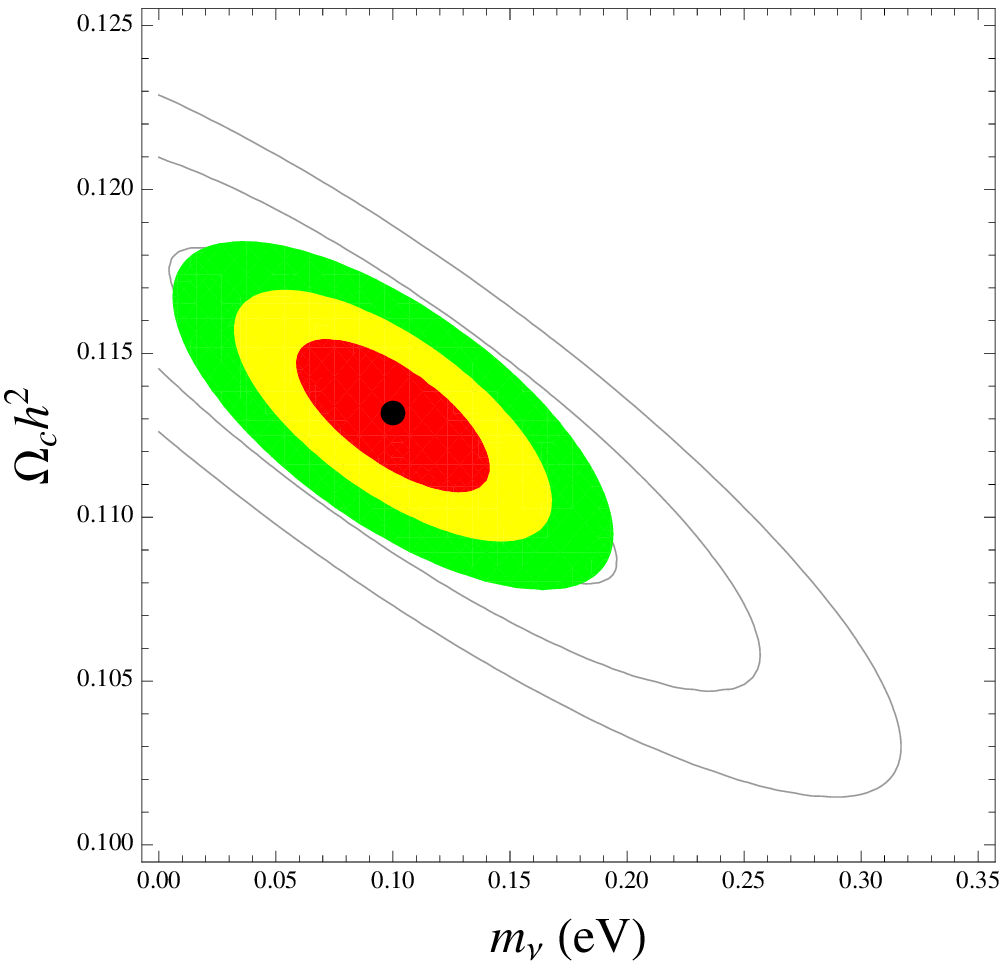}&
\includegraphics[width=4.3cm]{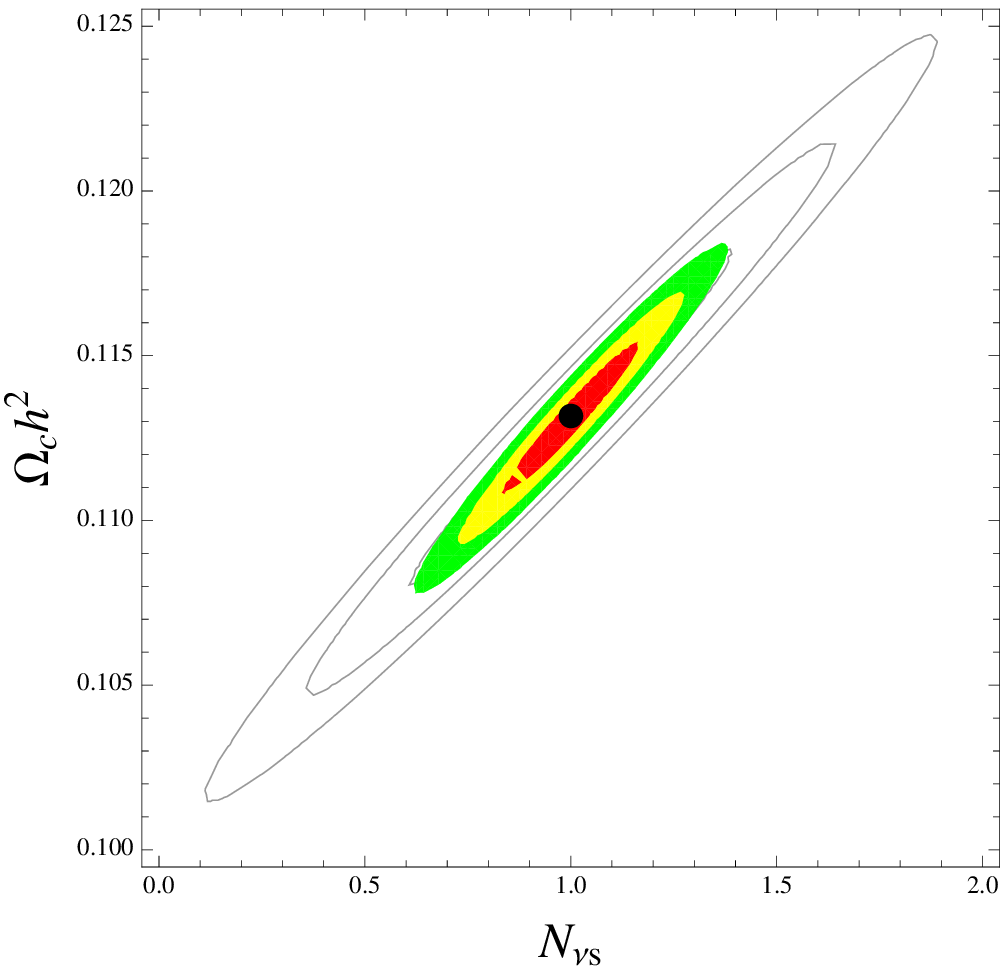}\\
\end{tabular}
\caption{The empty (filled) contours denote the $68\%$, $95\%$ and and $99.73\%$ CL regions for Planck plus BOSS (Euclid) data. The neutrino parameters in the fiducial model  are $N_{\nu_s}= 1$, $m_{\nu_s}=0.3$~eV and $m_\nu=0.1$~eV.}
\label{fig_omc}
\end{figure}

Figure \ref{fig_omc}, left panel, shows the correlation between the active neutrino mass $m_{\nu}$ and the cold dark matter energy density $\Omega_{c}h ^2$. Notice that the extraction of the cold dark matter component arise mostly from Planck CMB measurements. At late times, neutrinos contribute as an additional ingredient to the dark matter fluid and therefore a higher neutrino mass is compensated by a lower cold dark matter energy density. The right panel of Fig.~\ref{fig_omc} shows the correlation between cold dark matter and the sterile neutrino abundance. These two parameters are mostly extracted from CMB Planck data~\footnote{However, the addition of galaxy clustering measurements help in breaking degeneracies}. The sterile neutrinos considered here with $0.3$~eV masses are relativistic at decoupling. A higher number of relativistic species will shift to a later period the matter radiation equality era and also enhance the first CMB acoustic peak. These effects can be compensated with a higher cold dark matter energy density, as shown by the positive correlation among the two parameters.

\subsection{Cosmological parameter shifts}

In order to test the capabilities of future experiments to discriminate between
different theoretical models, regardless of their parameters, we follow here the
method of Ref.~\cite{kitching}.

The idea is the following: if the data is fitted assuming a model
$M'$ with $n'$ parameters, but the true underlying cosmology is a model $M$ characterized
by $n$ parameters (with $n>n'$ and the parameter space of $M$ including the model $M'$ as a subset),
the inferred values of the $n'$ parameters will be shifted from their true values to
compensate for the fact that the model used to fit the data is wrong. In the case illustrated here, $M$ will
be the model \emph{with} massive sterile neutrinos and $M'$ the one \emph{without} massive sterile neutrinos.
While the first $n'$ parameters are the same for both models, the remaining $n-n'=p$ parameters in the
enlarged model $M$ are accounting for the presence of massive sterile neutrinos, i.e. $m_{\nu_s}$ and $N_{\nu_s}$.
Assuming a gaussian likelihood, the shifts of the remaining $n'$ parameters is given by \cite{kitching}:
\begin{equation}
\delta\theta'_\alpha =
-(F'^{-1})_{\alpha\beta}G_{\beta\zeta}\delta\psi_\zeta \qquad
\alpha,\beta=1\ldots n', \zeta=n'+1\ldots n \label{offset}~,
\end{equation}
where $F'$ represents the Fisher sub-matrix for the model $M'$ without massive steriles and $G$ denotes the Fisher matrix for the model $M$ with $m_{\nu_s}, N_{\nu_s} > 0$.

We have computed the shifts induced in the cosmological parameters in several \emph{true} cosmologies with a number of sterile neutrinos $N_{\nu_s}=1, 2$ of masses $m_{\nu_s}=0.1, 0.3$ and $0.5$~eV. The mass of the active neutrino has been kept to $0.1$~eV. These cosmologies are then wrongly fitted to a cosmology without sterile massive neutrino species. While certain parameters are exclusively measured by CMB probes or by the 
combination of CMB and other cosmological data sets (like $\Omega_c h^2$, 
$\Omega_b h^2$, $n_s$ and $A_s$), there are other parameters such as the 
Hubble constant $H_0$ or the active neutrino mass $m_\nu$ which can 
be determined by other experiments. Then it is possible to verify 
the cosmological model assumptions by comparing the values 
of $H_0$ and $m_\nu$ extracted from CMB and LSS cosmological data to
the values of these parameters obtained by other experiments, as missions 
devoted to measure the Hubble constant and tritium beta decay experiments~\footnote{Neutrinoless double beta decay provides also a bound on the so-called effective neutrino mass $\langle m \rangle \equiv|\sum_i U^2_{ei} m_i|$. However, these bounds apply only in the case that neutrinos have a Majorana nature. Therefore, we focus on tritium beta decay constraints which apply regardless of the Dirac vs Majorana nature of the neutrino.}.
 The former experiments measure the electron neutrino mass $m_{\nu_e}$, which, in practice,
 when considering three active massive neutrinos, reads:
\begin{eqnarray}\label{eq:mnue}
m_{\nu_e}^2\equiv \sum_{i=1,3}  |U_{ei}^2| m_i^2~,
\end{eqnarray}
being $U_{ei}$ the first-row elements of the Pontecorvo-Maki-Nakagawa-Sakata leptonic mixing matrix. In the case of additional $N_{\nu_s}$ massive sterile neutrino species, $m_{\nu_e}$ would be given by
\begin{eqnarray}\label{eq:mnue}
m_{\nu_e}^2\equiv \sum_{i=1,3+N_{\nu_s}} |U_{ei}^2| m_i^2~,
\end{eqnarray}
Given the current best-fit values of the sterile-electron neutrino mixing terms $|U_{es}| < 10^{-1}$~\cite{Melchiorri:2008gq} and the sub-eV sterile neutrino masses considered here, we neglect the contribution of the sterile neutrino species to $m_{\nu_e}$. In the following, we apply the usual constraints on $m_{\nu_e}$ in our cosmological scenarios even if they contain massive sterile neutrino species. We therefore neglect the capability of beta decay experiments of measuring the individual neutrino masses and mixings. For a recent study of the KATRIN potential for sterile neutrino detection, see Ref.~\cite{hannestad}.

For sterile neutrino masses $m_{\nu_s}\sim 0.5$~eV and $N_{\nu_s}=1,2$, 
the shifts induced in $H_0$  are very large, for both BOSS and Euclid 
experiments combined with CMB Planck data. The reconstructed value of 
$H_0$ is within the range $\sim 20-50$~km/s/Mpc, values which are in strong
disagreement with current measurements of the Hubble parameter from 
HST~\cite{Riess:2009pu,Freedman:2000cf}. The reconstructed value of the 
active neutrino mass is also in some cases $m_\nu \sim 2$~eV which is 
the current $95\%$ CL limit from tritium $\beta$-decay 
experiments~\cite{Otten:2008zz}. Consequently, 
after combining near future BOSS and Planck data one would conclude that 
the cosmological model assumed with $m_{\nu_s}\sim 0.5$~eV and $N_{\nu_s}=1,2$ 
is wrong. The same situation will arise when $m_{\nu_s}\sim 0.3$~eV 
and two sterile massive species, $N_{\nu_s}=2$.

\begin{table}[htbp]
\begin{center}
\begin{tabular}{c|ccc}
\hline\hline
Parameter            &Fiducial      &Reconstructed     &Shift (\%) \\
\hline
$H_0$ [km/s/Mpc]       &70.5       & 50.5    &28\%  \\
$m_{\nu}$ [eV]         &0.30        & 0.98   &230\% \\
\hline\hline
\end{tabular}
\end{center}
\caption{Shifted values and relative changes for the parameters $H_0$ and $m_{\nu}$ when the true cosmology has $N_{\nu_s}=1$, $m_{\nu_s}=0.3$~eV and $m_{\nu}=0.1$~eV but BOSS plus Planck data are fitted to a cosmology with no sterile massive neutrino species.}
\label{tab:PB1}

\end{table}

\begin{table}[htbp]
\begin{center}
\begin{tabular}{c|ccc}
\hline\hline
Parameter       &Fiducial      &Reconstructed     &Shift (\%) \\
\hline
$H_0$ [km/s/Mpc]    &70.5      &65.0     &8\% \\
$m_{\nu}$ [eV]      &0.30      &0.48   &60\% \\
\hline\hline
\end{tabular}
\end{center}
\caption{Shifted values and relative changes for the parameters $H_0$ and $m_{\nu}$ when the true cosmology has $N_{\nu_s}=1$, $m_{\nu_s}=0.3$~eV and $m_{\nu}=0.1$~eV but Euclid plus Planck data are fitted to a cosmology with no sterile massive neutrino species.}
\label{tab:PE1}

\end{table}

For $m_{\nu_s}\sim 0.3$~eV and $N_{\nu_s}=1$, the shifts using both BOSS and 
Euclid data are reported in Tabs.~\ref{tab:PB1} and \ref{tab:PE1}. 
While the shift induced in the Hubble constant is very large for the BOSS case, for Euclid that shift is still consistent with current estimates of $H_0$. A number of experiments (HST, Spitzer, GAIA and JWST~\cite{Freedman:2010xv}) are expected to measure $H_0$ with $2\%$ uncertainty in the next decade and an inconsistency between the inferred $H_0$ values from these experiments and those from the cosmological probes considered here could point to the existence of additional sterile neutrino species. On the other hand, the aim of the tritium beta decay experiment KATRIN~\cite{Drexlin:2005zt} is a sensitivity of $m_{\nu_e}< 0.2$~eV at $90\%$ CL in case of a null result or a 5$\sigma$ discovery potential for $m_{\nu_e} \ge 0.35$~eV. Therefore, the reconstructed values of $m_\nu=0.48$~eV (Euclid plus Planck) and $0.98$~eV (BOSS plus Planck) could be easily testable by the KATRIN experiment. Similar results are obtained for smaller sterile neutrino masses $m_{\nu_s} \sim 0.1$~eV with a higher number of sterile species $N_{\nu_s}=2$.

For smaller sterile neutrino masses $m_{\nu_s} \sim 0.1$~eV and $N_{\nu_s}=1$, the shift induced
in $H_0$ is larger than $2\%$ for both BOSS and Euclid data (combined with Planck). Therefore
it would still be possible to check the fiducial cosmology with future measurements of $H_0$. The shift
induced on the active neutrino mass using Euclid data is negligible and this means that it would be possible
to recover the true value of the active neutrino mass even if the data is fitted to the wrong cosmology.
Thus, the combination
of Planck and Euclid data would not lead to an inconsistency between active neutrino mass estimates
from Planck and Euclid on the one hand, and beta decay experiments on the other hand. Regarding BOSS plus Planck data
however, the shift induced in the active neutrino mass
$m_\nu$ is of the order of $100\%$ and the comparison with an independent measurement of $m_\nu$ as that
performed by KATRIN could test the validity of the cosmological model assumptions.

We have shown above that if the true $N_{\nu_s}=1,2$, wrongfully assuming $N_{\nu_s}=0$ would lead to
discrepancies between the cosmological
probes considered here (large scale structure and CMB) and independent measurements of $H_0$ and $m_{\nu}$.
Of course, another clear
indicator that the assumed model is incorrect is simply that the $N_{\nu_s}=0$ would likely provide a bad fit
to the large scale structure and CMB data themselves.
However, the induced bias discussed above would provide a useful extra check when independent
measurements of $H_0$ and/or $m_{\nu}$ are available.
In addition, the bias calculation shows that even if one is not interested in the sterile neutrinos per se, not
taking them into account could lead to very wrong conclusions about the other cosmological parameters.

\section{Summary}
\label{sec:seciii}
Neutrino oscillation experiments have brought to light the first departure from the Standard Model of particle physics, indicating that neutrinos have non zero masses and opening the possibility for a number of extra sterile neutrinos. LSND and MiniBooNE antineutrino data require these extra sterile species to be massive. Much effort has been devoted in the literature to constrain the so called (3+1) (three active plus one sterile) and (3+2) (three active plus two sterile) models. 

Cosmology can set bounds on both the active and sterile neutrino masses as well as on the number of sterile neutrino species. We have explored here the current constraints on these parameters in the most natural scenario which corresponds to the case in which both the active and sterile neutrinos are massive particles. We find that models with two massive sub-eV sterile neutrinos plus three sub-eV active states are perfectly allowed at the $95\%$ CL by current Cosmic Microwave Background, galaxy clustering and Supernovae Ia data. The bounds derived here were obtained in the context of a $\Lambda$CDM cosmology and other scenarios with a dark energy component could allow for larger neutrino masses and/or abundances. We have also shown that Big Bang Nucleosynthesis Helium-4 and deuterium abundances exclude (3+2) models at the $95\%$ CL. However, the extra sterile states do not necessarily need to feature thermal abundances at decoupling. Their precise abundances are related to their mixings with the active neutrinos in the early universe.  

We have also forecasted the errors on the active and sterile neutrino parameters from Planck and galaxy survey
data. Future cosmological data are expected to measure sub-eV active and sterile neutrino masses and sterile
abundances with $10-30\%$ precision, for sub-eV ($0.5$~eV$>m_{\nu_s}>0.1$~eV) sterile neutrino masses. We have
also shown that the presence of massive sterile neutrinos in the universe could be inferred from inconsistencies
among the values of $H_0$ obtained from cosmic microwave and galaxy clustering probes and those arising from
independent measurements of the Hubble constant over the next decade.  The validity of the cosmological assumptions
could also be tested by comparing cosmological measurements of the active neutrino mass with those obtained from
tritium beta decay experiments.




\begin{thebibliography}{99}

\frenchspacing

\bibitem{GonzalezGarcia:2007ib}
  M.~C.~Gonzalez-Garcia and M.~Maltoni,
  Phys.\ Rept.\  {\bf 460}, 1 (2008)
  [arXiv:0704.1800 [hep-ph]].
\bibitem{Lesgourgues:2006nd}
  J.~Lesgourgues and S.~Pastor,
  Phys.\ Rept.\  {\bf 429}, 307 (2006)
  [arXiv:astro-ph/0603494].
\bibitem{Komatsu:2010fb}
  E.~Komatsu {\it et al.}  [WMAP Collaboration],
  Astrophys.\ J.\ Suppl.\  {\bf 192}, 18 (2011)
  [arXiv:1001.4538 [astro-ph.CO]].
\bibitem{Reid:2009nq}
  B.~A.~Reid, L.~Verde, R.~Jimenez and O.~Mena,
  JCAP {\bf 1001}, 003 (2010)
  [arXiv:0910.0008 [astro-ph.CO]].
\bibitem{Hamann:2010pw}
  J.~Hamann, S.~Hannestad, J.~Lesgourgues, C.~Rampf and Y.~Y.~Y.~Wong,
  JCAP {\bf 1007}, 022 (2010)
  [arXiv:1003.3999 [astro-ph.CO]].
\bibitem{Mangano:2006ur}
  G.~Mangano, A.~Melchiorri, O.~Mena, G.~Miele and A.~Slosar,
  JCAP {\bf 0703}, 006 (2007)
  [arXiv:astro-ph/0612150].
\bibitem{Hamann:2007pi}
  J.~Hamann, S.~Hannestad, G.~G.~Raffelt and Y.~Y.~Y.~Wong,
  JCAP {\bf 0708}, 021 (2007)
  [arXiv:0705.0440 [astro-ph]].
\bibitem{Mangano:2010ei}
  G.~Mangano, G.~Miele, S.~Pastor, O.~Pisanti and S.~Sarikas,
  JCAP {\bf 1103} (2011) 035
  [arXiv:1011.0916 [astro-ph.CO]].
 \bibitem{Aguilar:2001ty}
  A.~Aguilar {\it et al.}  [LSND Collaboration],
  Phys.\ Rev.\  D {\bf 64}, 112007 (2001)
  [arXiv:hep-ex/0104049].
\bibitem{Sorel:2003hf}
  M.~Sorel, J.~M.~Conrad and M.~Shaevitz,
  Phys.\ Rev.\  D {\bf 70}, 073004 (2004)
  [arXiv:hep-ph/0305255].
\bibitem{Karagiorgi:2006jf}
  G.~Karagiorgi, A.~Aguilar-Arevalo, J.~M.~Conrad, M.~H.~Shaevitz, 
  K.~Whisnant, M.~Sorel and V.~Barger,
  Phys.\ Rev.\  D {\bf 75}, 013011 (2007)
  [Erratum-ibid.\  D {\bf 80}, 099902 (2009)]
  [arXiv:hep-ph/0609177].
\bibitem{AguilarArevalo:2007it}
  A.~A.~Aguilar-Arevalo {\it et al.}  [The MiniBooNE Collaboration],
  Phys.\ Rev.\ Lett.\  {\bf 98}, 231801 (2007)
  [arXiv:0704.1500 [hep-ex]].
\bibitem{AguilarArevalo:2009xn}
  A.~A.~Aguilar-Arevalo {\it et al.}  [MiniBooNE Collaboration],
  Phys.\ Rev.\ Lett.\  {\bf 103}, 111801 (2009)
  [arXiv:0904.1958 [hep-ex]].
\bibitem{Karagiorgi:2009nb}
  G.~Karagiorgi, Z.~Djurcic, J.~M.~Conrad, M.~H.~Shaevitz and M.~Sorel,
  Phys.\ Rev.\  D {\bf 80}, 073001 (2009)
  [Erratum-ibid.\  D {\bf 81}, 039902 (2010)]
  [arXiv:0906.1997 [hep-ph]].
\bibitem{Akhmedov:2010vy}
  E.~Akhmedov and T.~Schwetz,
  JHEP {\bf 1010} (2010) 115
  [arXiv:1007.4171 [hep-ph]].
\bibitem{Melchiorri:2008gq}
  A.~Melchiorri, O.~Mena, S.~Palomares-Ruiz, S.~Pascoli, A.~Slosar and M.~Sorel,
  JCAP {\bf 0901}, 036 (2009)
  [arXiv:0810.5133 [hep-ph]].
\bibitem{Acero:2008rh}
  M.~A.~Acero and J.~Lesgourgues,
  Phys.\ Rev.\  D {\bf 79} (2009) 045026
  [arXiv:0812.2249 [astro-ph]].
\bibitem{rt}
  J.~Hamann, S.~Hannestad, G.~G.~Raffelt, I.~Tamborra and Y.~Y.~Y.~Wong,
  Phys.\ Rev.\ Lett.\  {\bf 105}, 181301 (2010)
  [arXiv:1006.5276 [hep-ph]].
\bibitem{Dodelson:2005tp}
  S.~Dodelson, A.~Melchiorri and A.~Slosar,
  Phys.\ Rev.\ Lett.\  {\bf 97}, 04301 (2006)
  [arXiv:astro-ph/0511500].
\bibitem{camb}
  A.~Lewis, A.~Challinor and A.~Lasenby,
  Astrophys.\ J.\  {\bf 538}, 473 (2000)
  [arXiv:astro-ph/9911177].
%
\bibitem{Lewis:2002ah}
  A.~Lewis and S.~Bridle,
  Phys.\ Rev.\  D {\bf 66}, 103511 (2002)
  [arXiv:astro-ph/0205436].
\bibitem{wmap7}
D.~Larson {\it et al.},
  Astrophys.\ J.\ Suppl.\  {\bf 192}, 16 (2011)
  [arXiv:1001.4635 [astro-ph.CO]].

\bibitem{beth}
  B.~A.~Reid {\it et al.},
  Mon.\ Not.\ Roy.\ Astron.\ Soc.\  {\bf 404}, 60 (2010)
  [arXiv:0907.1659 [astro-ph.CO]].

\bibitem{Riess:2009pu}
  A.~G.~Riess {\it et al.},
  Astrophys.\ J.\  {\bf 699}, 539 (2009)
  [arXiv:0905.0695 [astro-ph.CO]].
\bibitem{sn}
  R.~Amanullah {\it et al.},
  Astrophys.\ J.\  {\bf 716}, 712 (2010)
  [arXiv:1004.1711 [astro-ph.CO]].
\bibitem{aver}
 E.~Aver, K.~A.~Olive and E.~D.~Skillman,
  JCAP {\bf 1005}, 003 (2010)
  [arXiv:1001.5218 [astro-ph.CO]].
\bibitem{it}
  Y.~I.~Izotov and T.~X.~Thuan,
  Astrophys.\ J.\  {\bf 710}, L67 (2010)
  [arXiv:1001.4440 [astro-ph.CO]].
\bibitem{hamann2}
  J.~Hamann, J.~Lesgourgues, G.~Mangano,
  JCAP {\bf 0803 } (2008)  004.
  [arXiv:0712.2826 [astro-ph]].

\bibitem{pettini}
  M.~Pettini, B.~J.~Zych, M.~T.~Murphy, A.~Lewis and C.~C.~Steidel,
  arXiv:0805.0594 [astro-ph].

\bibitem{iocco}
  O.~Pisanti, A.~Cirillo, S.~Esposito, F.~Iocco, G.~Mangano, G.~Miele and P.~D.~Serpico,
  Comput.\ Phys.\ Commun.\  {\bf 178}, 956 (2008)
  [arXiv:0705.0290 [astro-ph]].
  
\bibitem{Melchiorri:2007cd}
  A.~Melchiorri, O.~Mena and A.~Slosar,
  Phys.\ Rev.\  D {\bf 76}, 041303 (2007)
  [arXiv:0705.2695 [astro-ph]].
\bibitem{prep}
M. Blennow {\it et al.}, in preparation.
\bibitem{Mention:2011rk}
  G.~Mention, M.~Fechner, T.~Lasserre, T.~A.~Mueller, D.~Lhuillier, M.~Cribier and A.~Letourneau,
  arXiv:1101.2755 [hep-ex].
\bibitem{Carbone:2010ik}
  C.~Carbone, L.~Verde, Y.~Wang and A.~Cimatti,
  JCAP {\bf 1103}, 030 (2011)
  [arXiv:1012.2868 [astro-ph.CO]].
\bibitem{Tegmark:1996bz}
  M.~Tegmark, A.~Taylor and A.~Heavens,
  Astrophys.\ J.\  {\bf 480}, 22 (1997)
  [arXiv:astro-ph/9603021].
\bibitem{Jungman:1995bz}
  G.~Jungman, M.~Kamionkowski, A.~Kosowsky and D.~N.~Spergel,
  Phys.\ Rev.\  D {\bf 54}, 1332 (1996)
  [arXiv:astro-ph/9512139].
\bibitem{Fisher:1935bi}
  R.~A.~Fisher,
  Annals Eugen.\  {\bf 6} (1935) 391.
\bibitem{Perotto:2006rj}
  L.~Perotto, J.~Lesgourgues, S.~Hannestad, H.~Tu and Y.~Y.~Y.~Wong,
  JCAP {\bf 0610} (2006) 013
  [arXiv:astro-ph/0606227].
\bibitem{Planck}
    [Planck Collaboration],
  [arXiv:astro-ph/0604069]
\bibitem{Licia05}
  L.~Verde, H.~Peiris and R.~Jimenez,
  JCAP {\bf 0601}, 019 (2006)
  [arXiv:astro-ph/0506036].
\bibitem{eisenstein}
  H.~J.~Seo and D.~J.~Eisenstein,
  Astrophys.\ J.\  {\bf 598}, 720 (2003)
  [arXiv:astro-ph/0307460].
\bibitem{boss}
  D.~J.~Eisenstein {\it et al.}  [SDSS Collaboration],
  arXiv:1101.1529 [astro-ph.IM].
\bibitem{Euclid1}
A.~Refregier {\it et al.},
  arXiv:astro-ph/0610062.
\bibitem{Euclid2}
 A.~Refregier, A.~Amara, T.~D.~Kitching, A.~Rassat, R.~Scaramella, J.~Weller and f.~t.~E.~Consortium,
  arXiv:1001.0061 [astro-ph.IM].
\bibitem{Hannestad:2005gj}
S.~Hannestad,
Phys.\ Rev.\ Lett.\  {\bf 95} (2005) 221301
[arXiv:astro-ph/0505551].
\bibitem{LaVacca:2008mh}
  G.~La Vacca, S.~A.~Bonometto and L.~P.~L.~Colombo,
  New Astron.\  {\bf 14}, 435 (2009)
  [arXiv:0810.0127 [astro-ph]].
\bibitem{Gavela:2009cy}
  M.~B.~Gavela, D.~Hernandez, L.~L.~Honorez, O.~Mena and S.~Rigolin,
  JCAP {\bf 0907}, 034 (2009)
  [Erratum-ibid.\  {\bf 1005}, E01 (2010)]
  [arXiv:0901.1611 [astro-ph]].
\bibitem{kitching}
  A.~F.~Heavens, T.~D.~Kitching and L.~Verde,
  Mon.\ Not.\ Roy.\ Astron.\ Soc.\  {\bf 380}, 1029 (2007)
  [arXiv:astro-ph/0703191].
\bibitem{hannestad}
 A.~S.~Riis and S.~Hannestad,
  arXiv:1008.1495 [astro-ph.CO].

\bibitem{Freedman:2000cf}
  W.~L.~Freedman {\it et al.}  [HST Collaboration],
  Astrophys.\ J.\  {\bf 553}, 47 (2001)
  [arXiv:astro-ph/0012376].
\bibitem{Otten:2008zz}
  E.~W.~Otten and C.~Weinheimer,
  Rept.\ Prog.\ Phys.\  {\bf 71}, 086201 (2008)
  [arXiv:0909.2104 [hep-ex]].
\bibitem{Freedman:2010xv}
  W.~L.~Freedman and B.~F.~Madore,
  arXiv:1004.1856 [astro-ph.CO].
\bibitem{Drexlin:2005zt}
  G.~Drexlin  [KATRIN Collaboration],
  Nucl.\ Phys.\ Proc.\ Suppl.\  {\bf 145} (2005) 263.


\end{thebibliography}
\end{document}